\documentstyle[epsf,a4,12pt]{article}

\begin{document}

\begin{titlepage}

\vspace*{2.truecm}

\centerline{\Large \bf 
         Monte Carlo simulations and numerical solutions}
\centerline{\Large \bf of short-time critical dynamics
\footnote
{Work supported in part by DFG 
under the project TR 300/3-1}}
\vskip 0.6truecm

\centerline{\bf B. Zheng}
\vskip 0.2truecm
\centerline{FB Physik, Univ. Halle, 06099 Halle, Germany}

\vskip 2.5truecm

\abstract{Recent progress in numerical study
of the short-time critical dynamics is briefly reviewed.
}

\vspace{0.5cm}

{\small PACS: 64.60.Ht, 02.70.Lq}

{\small KEYWORDS: critical dynamics, Monte Carlo simulations}

\end{titlepage}

\section {Short-time dynamic scaling}

For a critical dynamic system in the 
{\it long-time} regime,  it is well known
that there exists a universal dynamic scaling form.
This is more or less due to that the correlation time
 is divergent and 
the spatial correlation length is also very large
in the long-time regime.

{\it Is there any universal behavior in the 
macroscopic short-time regime
of the critical dynamic evolution?} The traditional answer is no.
However, this has been changed in recent years.
In the macroscopic short-time regime,
the spatial correlation length is also small
in the macroscopic sense.
However, the large 
correlation time induces a memory effect.
The memory effect is represented by a scaling form.

Let us consider that
a magnetic system initially at
high temperature and with a small magnetization
is suddenly quenched to the critical
temperature,
 then released to dynamic evolution of model A.
A universal dynamic scaling form, which 
sets in right after a
time scale $t_{mic}$ which is enough large in {\it microscopic}
sense but still very small in
{\it macroscopic} sense, has been derived
with an $\epsilon$-expansion \cite {jan89}.
Important is that
extra critical exponents should be introduced
to describe the dependence of the scaling behavior on the initial
conditions.
For example, for the $k$-th moment of the magnetization,
the finite size scaling form 
 is written as \cite {jan89}
\begin{equation}
M^{(k)}(t,\tau,L,m_{0})=b^{-k\beta/\nu}
M^{(k)}(b^{-z}t,b^{1/\nu}\tau,b^{-1} L,
b^{x_{0}}m_{0}).
\label{eint50}
\end{equation}
Here $\beta$, $\nu$ are the well known static critical exponents
 and $z$ is the dynamic exponent,
while the {\it new independent} exponent $x_0$ 
is the scaling dimension
of the initial magnetization $m_0$.

In this and the next sections we discuss
 the dynamics generated by Monte Carlo algorithms,
which belongs to model A \cite {hoh77}.

\begin{figure}
\epsfysize=5.5cm
\epsfclipoff
\fboxsep=0pt
\setlength{\unitlength}{1cm}
\begin{picture}(8,8)(0,0)
\put(-1,0){\epsffile{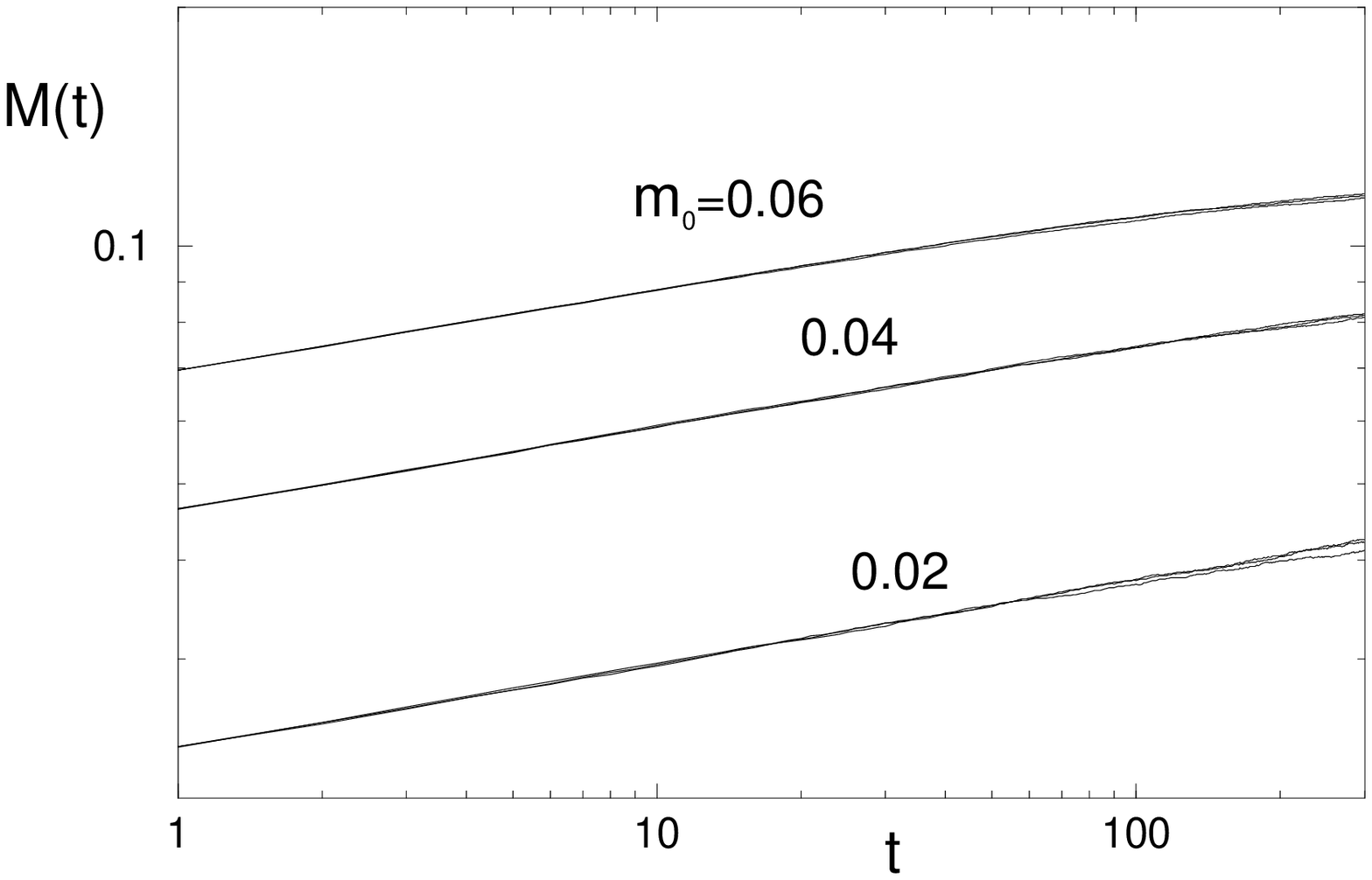}}
\epsfysize=5.5cm
\put(7,0){\epsffile{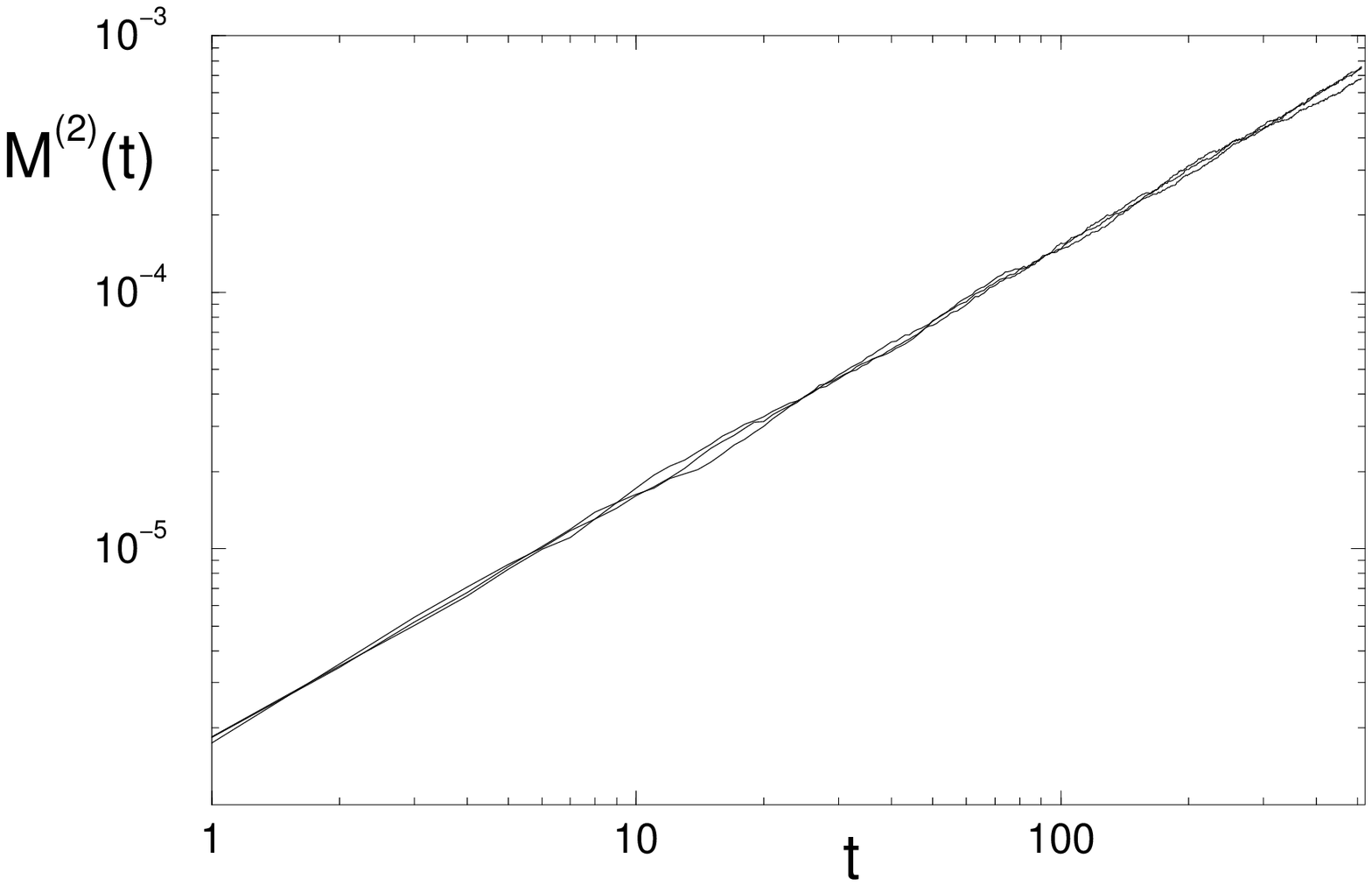}}
\end{picture}
\caption{ 
(a) The magnetization of the 3D Ising model
with $L=128$. (b) The second moment.
}
\label{f1}
\end{figure}

For a large lattice ($L=\infty$) and $\tau=0$ ,
 from the scaling form (\ref{eint50})
 one derives for small enough
 $t$ and $m_0$ \cite {jan89,zhe98}
\begin{equation}
M(t) \sim m_0 \, t^\theta, \ \ \ \ \theta =(x_0 - \beta /\nu)/z.
\end{equation}
For almost all statistical systems studied up to now,
the exponent $\theta$ is {\it positive},
i.e. the magnetization undergoes surprisingly
 {\it a critical initial increase} \cite {jan89,zhe98}.
 The time scale of this increase is $t_0 \sim m_0 ^ {-z/x_0}$.
 In Fig. \ref {f1} (a), the magnetization for the 3D Ising model
 with the heat-bath algorithm has been displayed in log-log scale.
 Power law behavior is observed 
 at beginning of the time evolution. The microscopic time scale 
 $t_{mic}$ is extremely small in this case. 
 Extrapolating the results to $m_0=0$ one obtains
 $\theta=0.108(2)$ \cite {jas99}.
     
Other two interesting observables in short-time dynamics 
are the auto-correlation and the second moment of the magnetization.
For $\tau=0$ and $m_0=0$, it is well known
\cite {hus89,zhe98}
\begin{equation}
M^{(2)}(t) \sim  t^{c_2}, \ \ \ \ c_2 =(d - 2\beta /\nu)/z.
\label{emag30}
\end{equation}
In Fig. \ref {f1} (b), the second moment for the 3D Ising model
with the heat-bath algorithm
is plotted in log-log scale \cite {jas99}.
 Almost perfect power law is seen.
Careful analysis reveals \cite {jan92}
the auto-correlation ($\tau=0$)
\begin{equation}
A(t)\sim t^{-c_a},\qquad c_a=\frac{d}{z}-\theta.
\label{ca}
\end{equation}
Interesting here is that even though $m_0=0$,
the exponent $\theta$ still enters the auto-correlation.
The behavior in Eq. (\ref {ca}) has been confirmed 
in many systems \cite {zhe98,hus89}.

In Table \ref {t1}, we have summarized the measured 
critical exponents of the 3D Ising model.
Taking the exponent $\theta$ as an input,
from $c_a$ we estimate the dynamic exponent $z$.
With $z$ in hand, from $c_2$ we obtain the static exponent
$2\beta/\nu$. The values of $z$ and $\beta/\nu$
agree nicely with those measured in equilibrium,
$z=2.04(3)$ \cite {wan91} and $2\beta/\nu=1.036(14)$
\cite {fer91}.
This fact, on the one hand, strongly supports
the short-time dynamic scaling, and on the other hand,
provides new ways for the numerical measurements 
of the critical exponents.

\begin{table}\centering
\begin{tabular}{ccccc}
    $\theta$  & $c_a$ &  $z$ &  $c_2$ &  $2\beta/\nu$   \\
\hline
    0.108(2)  & 1.36(1) & 2.04(2) & 0.970(11)  & 1.02(3)  \\
\end{tabular}
\caption{The exponents of the 3D Ising model.
}
\label{t1}
\end{table}

\section {Dynamic measurements of critical exponents}

For critical dynamics,
correlation times are very large.
Therefore, for standard Monte Carlo simulations in equilibrium
it is very difficult to generate independent
spin configurations. This is the so-called
{\it critical slowing down}.
One of the most succeesful methods to overcome critical slowing down
is the non-local cluster method.
However, it does not apply to
systems with quenched randomness or/and frustration.

In short-time dynamics, we do not have the problem
of generating independent spin configurations.
Therefore, one does not suffer from 
critical slowing down.
The 2D fully frustrated XY (FFXY) model is a typical example
where critical slowing down is severe.

In general, for determination of the dynamic exponent $z$ 
and static exponents a dynamic process starting from
a completely {\it ordered} state is more favorable,
since fluctuation is much less.
The completely ordered state in the 2D FFXY model is 
the ground state where spins
on four sublattices orient in different directions.
Here we only concern the chiral degrees of freedom
and the phase transition is of the second order. 
Assuming the lattice is sufficiently large,
around the critical point the dynamic scaling form
of the magnetization is written as
\begin{equation}
M(t,\tau)=t^{-\beta/\nu z}
F(t^{1/\nu z}\tau) \ .
\label{em1}
\end{equation}
If $\tau=0$, the magnetization decays by a power law 
$M(t) \sim t^{-\beta/\nu z}$. 
If $\tau \not= 0$, the power law behavior is modified by
the scaling function $F(t^{1/\nu z}\tau)$.
From this fact, one determines the critical point
and the critical exponent $\beta/\nu z$.

\begin{figure}[p]
\epsfysize=6.5cm
\epsfclipoff
\fboxsep=0pt
\setlength{\unitlength}{0.6cm}
\begin{picture}(13.6,12)(0,0)
\put(-1.,0.3){{\epsffile{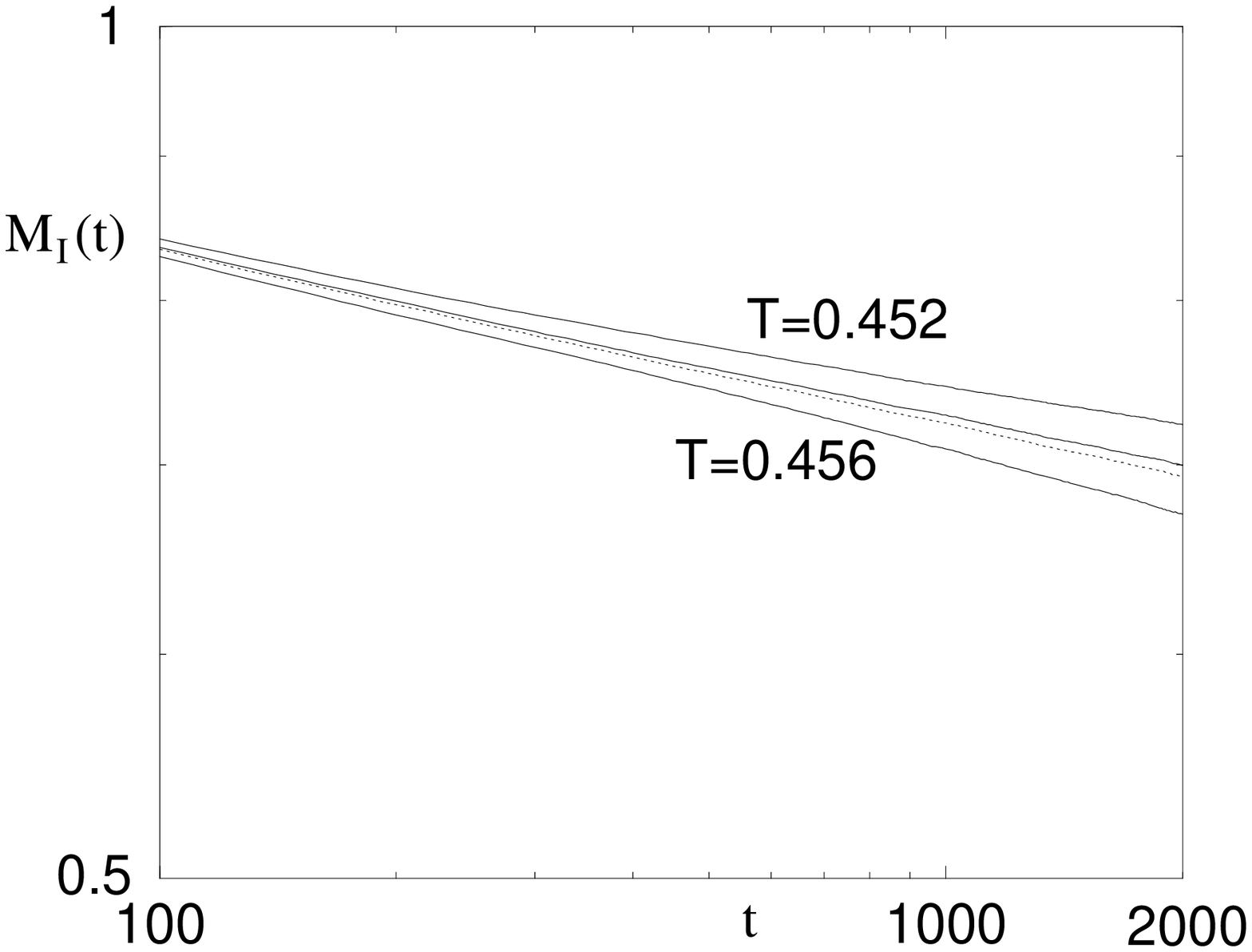}}}
\epsfysize=6.5cm
\put(12.,0.3){{\epsffile{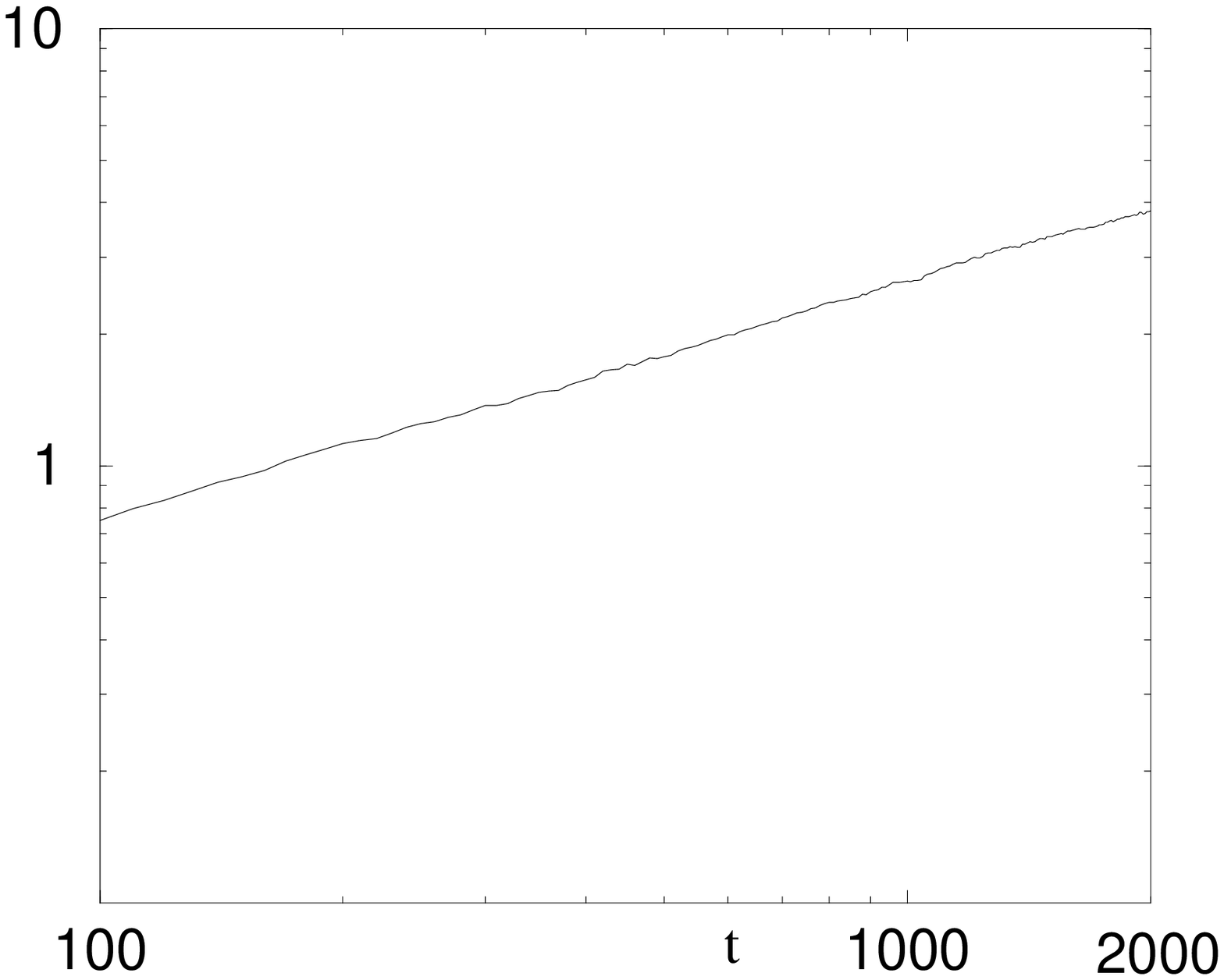}}}
\put(18.0,4.5){\makebox(0,0){\footnotesize $\partial_\tau \ln M_I$}}
\end{picture}
\caption{ (a)The chiral magnetizations 
of the 2D FFXY model
starting from an ordered state. From above, the solid lines
represent $M_I(t)$ at T=0.452, 0.454 and 0.456. The dotted line is
at $T_c=0.4545$.
(b) The derivative $\partial_\tau ln M_I(t, \tau)|_{\tau=0}$. }
\label{f3}
\end{figure}

In Fig. \ref {f3} (a), the magnetization at $T=0.452$,
$0.454$ and $0.456$ is plotted in double-log scale.
The Metropolis algorithm has been used in simulations.
Extrapolating
$M(t)$ to other temperatures,
from the time interval [200,2000] we measure
the critical temperature to be $T_c=0.4545(2)$ and  
$\beta / \nu z = 0.0602(2)$ \cite {luo98}.
To determine the exponent $1/(\nu z)$, we differentiate
$\ln \ M(t,\tau)$ and obtain
\begin{equation}
\partial_\tau \ \ln \ M(t,\tau)|_{\tau=0} \sim  t^{1/\nu z} \ . 
\label{em0140}
\end{equation}
This power law behavior is shown in Fig. \ref {f3} (b)
\cite {luo98}.

In order to estimate the dynamic exponent $z$ {\it independently},
we introduce a Binder cumulant 
$U = M^{(2)}/M^2 -1$, 
and finite size scaling analysis shows
\begin{equation}
U(t,L) \sim t^{d/z}.
\label{em0160}
\end{equation}
Now we complete the measurements of the critical exponents.
The results are given in Table \ref {t2},
in comparison to those obtained in equilibrium simulations.
Our results obviously support that the exponent $\nu$
of the 2D FFXY model is different from that of the 2D Ising model.
But other exponents of both models look similar.
\begin{table}[h]\centering
$$
\begin{array}{|c|c|c|c|c|c|c|}
\hline
       & Ref. \cite {luo98} & Ref. \cite {jos96} & Ref. \cite {ols95}
         & Ref. \cite {lee94} & Ref. \cite {gra93} & Ising\\
       &(1998)&(1996)&(1995)&(1994)&(1993) &\\
\hline
 T_c   & 0.4545(2) &0.451(1) &0.452(1)& 0.454(2) & 0.454(3) &\\
\hline
 \nu &0.81(2) & 0.898(3) &1  &0.813(5) &0.80(5) & 1\\
\hline
2\beta /\nu &0.261(5)&&&0.22(2)&0.38(2) & 0.25 \\
\hline
z &2.17(4)&&&& & 2.165(10)\\ 
\hline
\theta &0.202(3)&&&& & 0.191(3)\\ 
\hline
\end{array}
$$
\caption{
Dynamic measurements of the exponents of the 2D FFXY model
compared with those in recent references.
 For the Ising model, 
 the exponent $\theta$ is taken from 
Refs. \protect\cite {gra95}. The exponent $z$ 
in literature ranges from $2.155$ to $2.172$ 
\protect\cite {zhe98}.
Here an `average' value is given.}
\label{t2}
\end{table}

\section {Deterministic dynamics}

Up to now, we have discussed {\it stochastic} dynamics
induced by Monte Carlo algorithms.
However, in general stochastic dynamics is $not$
equivalent to real dynamics described 
by microscopic {\it deterministic} equations of motion.
Only in some cases stochastic dynamics is an effective description
 of real physical systems.
Therefore, it is important and interesting to investigate
whether there exists universal scaling behavior 
in deterministic dynamics.

We consider the 2D $\phi^4$ theory.
The Hamiltonian on a square lattice is
\begin{equation}
H=\sum_i \left [ \frac{\pi_i^2}{2}  
  + \frac{1}{2} \sum_\mu (\phi_{i+\mu}-\phi_i)^2 
  - \frac{m^2}{2} \phi_i^2 
  + \frac{\lambda}{4!} \phi_i^4 \right ]
\label{e30}
\end{equation}
with $\pi_i=\dot \phi_i$ and it leads to the equations of motion
\begin{equation}
 \ddot \phi_i= \sum_\mu (\phi_{i+\mu}+\phi_{i-\mu}- 2\phi_i)
  +  m^2 \phi_i
  - \frac{\lambda}{3!} \phi_i^3\ .
\label{e40}
\end{equation}

In dynamic evolution governed by
 Eq. (\ref {e40}), energy is conserved.
The solutions are assumed to generate a microcanonical ensemble.
The temperature could be defined as the averaged
kinetic energy. In the short-time dynamic approach, however,
the total energy is a more convenient controlling parameter,
 since it is conserved and 
can be input from the initial state. 
Therefore, from now $\tau$ 
will be understood as a reduced energy density 
$(\epsilon-\epsilon_c)/\epsilon_c$.
Here $\epsilon_c$ is the critical energy density.

The order parameter of the $\phi^4$ theory is the magnetization.
For the dynamic process starting from a {\it random} state
\cite {zhe99},
we assume a same dynamic scaling form as in stochastic dynamics.
For sufficiently large $L$ and small enough $m_0$ and $t$,
from scaling form (\ref {eint50}) one deduces 
\begin{equation}
M(t)=m_0t^{\theta}F(t^{1/\nu z} \tau).
\label{e90}
\end{equation}
Similar as the discussion on Eq. (\ref {em1}) in Sec. 2,
we determine the critical energy $\epsilon_c$
and the exponent $\theta$ from Eq. (\ref {e90})
and the exponent $1/\nu z$ from its derivative.
Finally, the exponent $\beta/\nu z$ and the dynamic
exponent $z$ are estimated from
the second moment and the auto-correlation.

\begin{figure}[p]\centering
\epsfysize=6.cm
\epsfclipoff
\fboxsep=0pt
\setlength{\unitlength}{0.6cm}
\begin{picture}(9,9)(0,0)
\put(-2,-0.5){{\epsffile{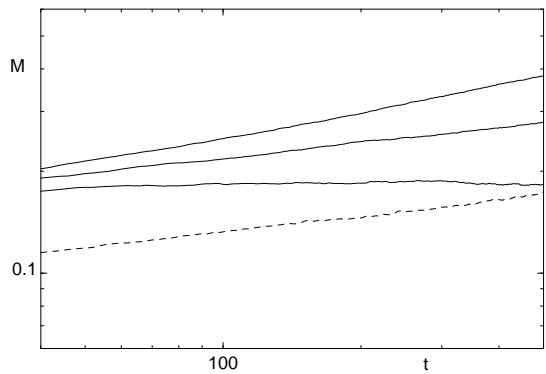}}}
\end{picture}
\caption{The magnetization of the $\phi^4$ theory in log-log scale. 
Solid lines are for $m_0=0.015$ with energy densities
$\epsilon=20.7$, $21.1$ and $21.5$ (from above),
while the dashed line is for $m_0=0.009$ with $\epsilon_c=21.11$.
}
\label{f4}
\end{figure}

In Fig.~\ref {f4}, numerical solutions of the magnetization
with $m_0=0.015$
for parameters  $m^2=2.$
and $\lambda=0.6$
have been plotted with solid lines for three energy densities
$\epsilon=20.7$, $21.1$ and $21.5$
in log-log scale. Careful analysis of the data between $t=50$ and $500$
leads to the critical energy density
 $\epsilon_c=21.11(3)$ \cite {zhe99}.
 This agrees well with $\epsilon_c=21.1$ given 
 by the Binder cumulant in
 equilibrium in Ref.~\cite {cai98}.
 At $\epsilon_c$, one measures the exponent $\theta=0.146(3)$.
The magnetization with $m_0=0.009$
 is also displayed in Fig.~\ref {f4}
 (dashed line). The corresponding exponent is $\theta=0.158(2)$.
 Extrapolating the results
 to $m_0=0$, we obtain the final value $\theta=0.176(7)$.
With the critical energy in hand, 
we estimate other exponents.
In Table~\ref {t3}, we summarize all 
the critical exponents of the $\phi^4$ theory.
Remarkably, not only the static exponents, but also
the dynamic exponents  $z$ of the $\phi^4$ theory
is the same as those of the Ising model
 with standard Monte Carlo dynamics.

\begin{table}[h]\centering
\begin{tabular}{cccccccc}
       & $\theta$  & $d/z-\theta$ & $(d-2\beta/\nu)/z$ & $1/\nu z$
                 & $z$          & $2\beta/\nu$       & $\nu$     \\
\hline
$\phi^4$ & 0.176(7)& 0.755(5)   & 0.819(12)        & 0.492(26)
                 & 2.148(20)  & 0.24(3)        &  0.95(5) \\
Ising  & 0.191(1)& 0.737(1)   & 0.817(7)         &
                 & 2.155(3)   & 1/4              & 1\\
\end{tabular}
\caption{The critical exponents of the $\phi^4$ theory
in comparison with those of the Ising model
Ref.~\protect\cite {zhe98}.
}
\label{t3}
\end{table}

\section {Concluding remarks}

We have demonstrated that universal scaling behavior
emerges already in the macroscopic short-time regime
of the critical dynamic evolution.
The short-time dynamic scaling is found not only in stochastic dynamics
but also in microscopic deterministic dynamics.
Therefore, it is fundamental.
Furthermore, it leads to new methods for numerical measurements
of both dynamic and static critical exponents.
The methods do not suffer from critical slowing down. 

%\bibliographystyle{stybase/pr_np}
%\bibliography{stybase/ising}

\end{document}